\documentclass[12pt]{iopart}

\begin{document}

\newcommand{\be}{\begin{equation}}
\newcommand{\ee}{  \end{equation}}
\newcommand{\ba}{\begin{eqnarray}}
\newcommand{\ea}{  \end{eqnarray}}
\newcommand{\ve}{\varepsilon}

\title[Spectral fluctuation properties]{Spectral fluctuation
properties of constrained unitary ensembles of Gaussian--distributed
random matrices}

\author{Z Pluha\v r$^1$ and H A Weidenm{\"u}ller$^2$}

\address{$^1$ Institute of Particle and Nuclear Physics, Faculty of
Mathematics and Physics, Charles University, 18000 Praha 8, Czech
Republic}

\address{$^2$ Max-Planck-Institut f{\"u}r Kernphysik, Postfach 103980,
69029 Heidelberg, Germany}

\ead{Hans.Weidenmueller@mpi-hd.mpg.de}

\begin{abstract}
We investigate the spectral fluctuation properties of constrained
ensembles of random matrices (defined by the condition that a number
$N_Q$ of matrix elements vanish identically; that condition is imposed
in unitarily invariant form) in the limit of large matrix dimension.
We show that as long as $N_Q$ is smaller than a critical value (at
which the quadratic level repulsion of the Gaussian unitary ensemble
of random matrices may be destroyed) all spectral fluctuation measures
have the same form as for the Gaussian unitary ensemble.
\end{abstract}

\pacs{05.45.-a, 02.50.Ey, 24.60.Lz}

\submitto{\JPA}

\section{Introduction}
\label{int}

We investigate the spectral fluctuation properties of the constrained
unitary ensembles of Gaussian--distributed random matrices (CGUE)
introduced in Ref.~\cite{Pap06}. Constrained ensembles of random
matrices deserve interest because they represent entire classes of
non--canonical random--matrix ensembles that were proposed to mimic
typical properties of interacting many--fermion systems more closely
than do the canonical ensembles (the Gaussian Orthogonal, Unitary, and
Symplectic Ensembles)~\cite{Meh91}. Examples for constrained ensembles
are the Embedded Gaussian Orthogonal Ensemble~\cite{Mon75,Ben03} and
the Two--Body Random Ensemble~\cite{Fre70,Boh71,Pap07}. The
constraints essentially require certain matrix elements or linear
combinations of matrix elements to vanish. It is difficult to deal
with the constrained ensembles analytically because they lack the
invariance properties that give analytical access to the canonical
ensembles. That difficulty is overcome by imposing the condition of
unitary, orthogonal, or symplectic invariance on the constrained
ensembles. It is expected that that condition leaves the spectral
fluctuations unchanged (whereas the eigenfunctions acquire the same
distribution as for the canonical ensembles). Here we focus on the
case of unitary invariance, i.e., on the CGUE.

Some spectral properties of the CGUE were exhibited in
Ref.~\cite{Pap06}. In particular, the following sufficient condition
for level repulsion was given. The quadratic level repulsion
characteristic of the GUE (the ensemble of Gaussian--distributed
unitary random matrices) also prevails for the CGUE provided that the
number $N_Q$ of constraints does not exceed a critical value,
\be
N_Q < N^{\rm crit}_Q
\label{1}
\ee
with $N^{\rm crit}_Q$ defined in Eq.~(\ref{A1}) below.

In the present paper we go beyond Ref.~\cite{Pap06} and address the
spectral fluctuation properties of the CGUE in the limit of large
matrix dimension $N \gg 1$. We do so for $N_Q < N^{\rm crit}_Q$. For
$N_Q = 0$ the CGUE coincides with the GUE. For $N_Q \geq N^{\rm
crit}_Q$ the form of the constraints does not seem to permit
definitive analytical statements. It has remained an open question to
what extent the spectral fluctuation properties of the CGUE (beyond
the statement of sheer level repulsion) are the same or differ from
those of the GUE for $0 < N_Q < N^{\rm crit}_Q$. We prove the
following

\noindent
{\bf Theorem}: For matrix dimension $N \gg 1$ and the number of
constraints $N_Q <N^{\rm crit}_Q$ (with $N^{\rm crit}_Q$ defined in
Eq.~(\ref{A1}) below), the spectral fluctuation measures of the CGUE
coincide with those of the GUE save for correction terms of order $1 /
N$.

To make the paper self--contained, we collect in Section~\ref{def}
some definitions and results given in Ref.~\cite{Pap06}. In
Section~\ref{mod} we slightly modify the definition of the constrained
ensembles so as to remove a singularity. In Section~\ref{asym} we
discuss the form of the constraints in the limit $N \gg 1$. Our proof
is given in Section~\ref{proof}. It is based on an approach developed
in Ref.~\cite{Hac95}. Section~\ref{dis} contains a discussion. Some
technical details are presented in an Appendix.

\section{Definitions}
\label{def}

We consider Hermitean matrices acting on a Hilbert space ${\cal H}$
of dimension $N$. For any two such matrices $A$ and $B$, we introduce
the canonical scalar product in terms of the trace
\be
\langle A | B \rangle \equiv {\rm Tr} (A B) \ .
\label{scalar}
\ee
This allows us to define an orthonormal basis of $N^2$ Hermitean
basis matrices $B_\alpha = B^{\dagger}_\alpha$ which obey
\be
\label{basis}
\langle B_\alpha|B_\beta\rangle \equiv {\rm Tr} (B_\alpha B_\beta) 
= \delta_{\alpha \beta}
\label{orth}
\ee
and
\be
\sum_{\alpha = 1}^{N^2} | B_\alpha \rangle \langle B_\alpha | =
{\bf 1}_N
\label{compl}
\ee
where ${\bf 1}_N$ is the unit matrix in $N$ dimensions. Any Hermitean
matrix $H$ acting on ${\cal H}$ can be expanded in terms of the $N^2$
Hermitean basis matrices $B_\alpha$ as
\be
\label{expan}
H=\sum_{\alpha=1}^{N^2} h_\alpha B_\alpha.
\ee
The Gaussian Unitary Ensemble (GUE) of random matrices is obtained by
assuming that the expansion coefficients $h_\alpha$ are uncorrelated
Gaussian--distributed real random variables with mean value zero and a
common variance. For the GUE, the probability density W(H) of the
matrix elements of $H$ has the form
\be
W(H){\rm d} [H] = {\cal N}^{\rm GUE} \exp \bigg( - \frac{N}{2
\lambda^2} \langle H | H \rangle \bigg) {\rm d} [H] \ .
\label{GUE}
\ee
Here ${\rm d} [H]$ is the product of the differentials of all
independent matrix elements, ${\cal N}^{\rm GUE}$ is a normalization
factor, and $2 \lambda$ is the radius of Wigner's semicircle.

We introduce the constraints by considering two orthogonal subspaces
labeled ${\cal P}$ and ${\cal Q}$ with dimensions $N_P$ and $N_Q = N^2
- N_P$, respectively. These are defined in terms of orthogonal
projection operators
\ba
{\cal P} &=& \sum_{p =1}^{N_P} | B_p \rangle \langle B_p
| \ , \nonumber \\
{\cal Q} &=& \sum_{q = N_P + 1}^{N^2} | B_q \rangle \langle B_q
| \ .
\label{A3}
\ea
We have
\be {\cal P}^{\dagger} = {\cal P} \ , \ {\cal Q}^{\dagger} =
{\cal Q} \ , \ {\cal P}^2 = {\cal P} \ , \ {\cal Q}^2 = {\cal Q} \
, \ {\cal P} {\cal Q} = 0 \ , \ {\cal P} + {\cal Q} = {\bf 1}_N \ .
\label{A4}
\ee
Constraints can be formulated in the form $\langle H | {\cal Q}
\rangle = 0$ for all $H$. In the CGUE such constraints are used in
unitarily invariant form. The CGUE is defined by the probability
density $W_{\cal P}$ of the matrix elements of $H$ given by \ba
W_{\cal P}(H) {\rm d} [ H ] &=& {\cal N}^{\rm GUE} \exp{\left( -
\frac{N} {2 \lambda^2} \langle H | H\rangle \right)} {\rm d} [ H ]
\nonumber \\ && \times \int {\rm d} [U] \bigg( \prod_q \ \delta (
\sqrt{\frac{N}{2 \pi \lambda^2}} \langle U B_q U^{\dagger} | H \rangle
) \bigg) \ .
\label{A7b}
\ea
The integral ${\rm d}[U]$ extends over the unitary group in $N$
dimensions. The Haar measure of the unitary group is normalized to
one, i.e.,
\be
\int {\rm d} [U] = 1 \ .
\ee
We diagonalize the matrix $H$ with the help of a unitary matrix $V$,
\be
H = V x V^{\dagger} \ ,
\label{A8}
\ee
where $x={\rm diag}(x_1,\ldots,x_N)$ is the diagonal matrix of the
eigenvalues. The integration measure becomes
\be
{\rm d} [H] \propto \Delta^2(x) {\rm d} [x] {\rm d} [V] \ ,
\label{dh}
\ee
where ${\rm d}x$ is the product of the differentials of the $N$
eigenvalues, where ${\rm d} [V]$ is the Haar measure of the unitary
group in $N$ dimensions, and where $\Delta(x)$ denotes the Vandermonde
determinant
\be
\Delta(x) = \prod_{1 \le \mu < \nu \le N} (x_\mu - x_\nu) \ .
\label{van}
\ee
Eq.~(\ref{dh}) shows that eigenvalues and eigenvectors of the CGUE are
uncorrelated random variables. The joint probability distribution
$P_{\cal P}(x)$ of the eigenvalues is given by
\be
P_{\cal P}(x) = {\cal N}_0 \exp{\left( - \frac{N}{2 \lambda^2} \langle
x | x \rangle \right)} \Delta^2(x) F_{\cal P}(H) \ ,
\label{A8a} 
\ee
where
\ba
F_{\cal P}(H) &\equiv& \int {\rm d} [ U ] \bigg( \prod_q \delta (
\sqrt{ \frac{N}{2 \pi \lambda^2} } \langle B_q | U H U^{\dagger}
\rangle ) \bigg) \nonumber \\
&=& \int {\rm d} [ U ] \bigg( \prod_q \delta ( \sqrt{ \frac{N}{2 \pi
\lambda^2} } \langle B_q | U x U^{\dagger} \rangle ) \bigg)
\label{A9a}
\ea
is a function of the eigenvalues $\{ x_\mu \}$ only, and where ${\cal
N}_0$ is another irrelevant normalization factor.  Comparison of
Eq.~(\ref{A7b}) and Eq.~(\ref{GUE}) shows that the eigenvalue
distribution of the CGUE differs from that of the GUE by the factor
$F_{\cal P}(H)$. GUE--type level repulsion is contained in the factor
$\Delta^2(x)$ in Eq.~(\ref{A8a}), and such level repulsion will
prevail also in the CGUE unless $F_{\cal P}(H)$ is singular whenever
two eigenvalues coincide. In Ref.~\cite{Pap06} it was shown that
$F_{\cal P}(H)$ cannot be singular if the number $N_Q$ of constraints
obeys the inequality
\be
N_Q < N^{\rm crit}_Q = N ( N - 1 ) / 2 - \sum_{j = 1}^J L_j ( L_j - 1
) / 2\ .
\label{A1}
\ee
Here it is assumed that the matrix $B = \sum_q s_q B_q$ with real
coefficients $s_q$ possesses asymptotically (all $s_q$ large) $J$ sets
of degenerate eigenvalues with multiplicities $L_j$, $j = 1,
\ldots, J$.

\section{Modified Form of the Constraints}
\label{mod}

The function $F_{\cal P}(H)$ embodies the constraints. Therefore, it
is the central object of study in this paper. The treatment of
$F_{\cal P}(H)$ simplifies when all confining matrices $B_q$ are
traceless. We believe that that case is physically the more
interesting one, for the following reason. We show in the Appendix
that whenever the $B_q$s are not traceless, there always exists an
orthogonal transformation of the set $\{B_q\}$ to a new set
$\{\tilde{B}_q\}$ such that $F_{\cal P}(H)$ is unchanged and that all
$\tilde{B}_q$ with $q > N_P + 1$ are traceless. The one constraining
matrix $\tilde{B}_{N_P + 1}$ that is not traceless, is the sum of a
traceless part and of a multiple of the unit matrix. But constraining
$H$ with a multiple of the unit matrix means that we constrain the
centroid of the spectrum of $H$. We cannot think of a physically
interesting situation where such a constraint would be
meaningful. This is why we focus attention on the case where all $B_q$
are traceless,
\be
\langle B_q \rangle = 0 \ {\rm for \ all} \ q = N_P + 1, N_P + 2,
\ldots, N^2 \ , 
\label{traceless}
\ee
and treat the more general case where the condition~(\ref{traceless})
is violated, in the Appendix. Here and in the sequel we use the symbol
$\langle A \rangle$ to denote the trace of the matrix $A$. This is
consistent with the definition~(\ref{scalar}).

For the developments in Section~\ref{proof} we note the following
properties of $F_{\cal P}(H)$. The function $F_{\cal P}(H)$ is real
(this follows from Eq.~(\ref{A9a})) and positive definite (this is
seen when we write the defining delta functions as limits of
Gaussians). Using Fourier transformation, we can write $F_{\cal P}(H)$
in Eq.~(\ref{A9a}) as an $N_Q$--fold Fourier integral,
\be
F_{\cal P}(H) = \bigg( \frac{\lambda^2}{2 \pi N} \bigg)^{N_Q / 2}
\prod_{q = 1}^{N_Q} \int {\rm d} s_q \int {\rm d} [U] \ \exp \{ i
\langle B(s) | U H U^\dag \rangle \} 
\label{F1}
\ee
where
\be
B(s) = \sum_q s_q B_q
\label{F2}
\ee
and where $s$ stands for the set $\{s_1, \ldots, s_{N_Q} \}$. The
integral over the unitary group can be worked out and with ${\rm d} s
= \prod_q {\rm d} s_q$ yields (see Ref.~\cite{Pap06})
\be
F_{\cal P}(H) \propto \bigg( \frac{\lambda^2}{2 \pi N} \bigg)^{N_Q
/ 2} \int {\rm d} s \ \frac{\det \exp \{ i x_\mu b_\nu(s) \} }{
\Delta(x) \Delta(b(s))} \ .
\label{F6}
\ee
Here the $b_\nu(s)$ are the eigenvalues of the matrix $B(s)$, and
$\Delta (b)$ is the Vandermonde determinant of the $b_\nu(s)$, see
Eq.~(\ref{van}). In Ref.~\cite{Pap06} it was shown that the integrals
over $s$ converge if the condition~(\ref{A1}) is met. Moreover,
inspection of Eq.~(\ref{F6}) shows that $F_{\cal P}(H)$ is not
singular when two eigenvalues $x_\mu, x_\nu$ coincide.

However, because of the form of the constraints (Eq.~(\ref{A9a})), the
function $F_{\cal P}(H)$ is singular when all eigenvalues of $H$
coincide. To see this we define
\be
\tilde{H} = H - \frac{{\bf 1}_N}{N} \langle H \rangle \ ,
\label{F6a}
\ee
use the assumption~(\ref{traceless}) and rewrite Eq.~(\ref{F1}) in
the form
\be
F_{\cal P}(H) = F_{\cal P}(\tilde{H}) = \bigg( \frac{\lambda^2}{2 \pi
N} \bigg)^{N_Q / 2} \int {\rm d} s \int {\rm d} [U] \ \exp \{ i \langle
B(s) | U \tilde{H} U^\dag \rangle \} \ .
\label{F4}
\ee
Eq.~(\ref{F4}) shows that $F_{\cal P}(\tilde{H})$ is singular when
$\tilde{H} = 0$, i.e., when all eigenvalues of $H$ coincide. The
singularity mirrors a singularity in the definition~(\ref{A9a}) of
$F_{\cal P}(H)$. Indeed, when $x_\mu = x_\nu = y$ for all $\mu, \nu =
1, \ldots, N$, the Dirac deltas in the definition~(\ref{A9a}) take the
form $\delta(y \langle B_q \rangle)$.  Because of
Eq.~(\ref{traceless}) each of these terms is singular. We avoid the
singularity by modifying the definition of $F_{\cal
P}(\tilde{H})$. Instead of $F_{\cal P}(\tilde{H})$, we consider the
constraining function
\be
\tilde{F}_{\cal P}(\tilde{H}) = \bigg( \frac{\langle \tilde{H}^2
\rangle}{N \lambda^2} \bigg)^{N_Q / 2} F_{\cal P}(\tilde{H}) \ .
\label{F5}
\ee
The factor in front of $F_{\cal P}(\tilde{H})$ guarantees that
$\tilde{F}_{\cal P}(\tilde{H})$ is not singular at $\tilde{H} = 0$. At
the same time, that factor is a function of the sum of the eigenvalues
$x_\mu$ only. Thus, that factor cannot modify the correlations of
close--lying eigenvalues $x_\mu$, and the spectral fluctuation
properties of the constrained ensembles defined by the constraining
functions $F_{\cal P}(\tilde{H})$ and $\tilde{F}_{\cal P}(\tilde{H})$
are the same. Moreover, for real $x_\mu$ the function $\tilde{F}_{\cal
P}(\tilde{H})$ is real and positive definite.  According to
Eq.~(\ref{F6}) $\tilde{F}_{\cal P}(\tilde{H})$ is not singular for
finite values of the $x_\mu$. Inspection shows that when one of the
eigenvalues, $x_\mu$ say, tends to infinity, $\tilde{F}_{\cal
P}(\tilde{H})$ cannot grow more strongly than some power of
$x_\mu$. That growth is much weaker than the Gaussian suppression of
large eigenvalues in Eq.~(\ref{A7b}). Hence the confinement of the
spectrum to a finite interval characteristic of the GUE persists also
for the CGUE with constraining function $\tilde{F}_{\cal
P}(\tilde{H})$ although the shape of the average spectrum may be
modified.

Collecting everything, we have
\be
\tilde{F}_{\cal P}(\tilde{H}) = \bigg( \frac{\langle (\tilde{H})^2
\rangle }{2 \pi N^2} \bigg)^{N_Q / 2} \int {\rm d} [U] \int {\rm d}
s \ \exp \{ i \langle U B(s) U^\dag | \tilde{H} \rangle \} \ .
\label{F7}
\ee
It is convenient to introduce the new variables $t_q = \lambda s_q$.
Then
\be
\tilde{F}_{\cal P}(\tilde{H}) = \bigg( \frac{\langle (\tilde{H})^2
\rangle}{2 \pi \lambda^2 N^2} \bigg)^{N_Q / 2} \int {\rm d} [U] \int
{\rm d} t \ \exp \{ i \langle U B(t) U^\dag | (\tilde{H} / \lambda)
\rangle \} \ .
\label{F7a}
\ee
This shows that $\tilde{F}_{\cal P}(\tilde{H})$ depends on $\tilde{H}$
only via the dimensionless ratio $\tilde{H} / \lambda$, as expected.
Because of unitary invariance, $\tilde{F}_{\cal P}(\tilde{H})$ can
depend only on unitary invariants constructed from $\tilde{H} /
\lambda$. The only such invariants are the normalized traces of
$\tilde{H}^n / \lambda^n$ with positive integer $n$. For $N \gg 1$
this is shown explicitly in the next Section. The probability density
for the Hamiltonian matrices of the CGUE is given by
\be
\tilde{W}_{\cal P}(H) {\rm d} [ H ] = \tilde{\cal N} \exp{ \left( -
\frac{N} {2 \lambda^2} \langle H | H\rangle \right)} \tilde{F}_{\cal
P}(\tilde{H}) \ {\rm d} [ H ] \ .
\label{F8}
\ee
The substitution of $F_{\cal P}(\tilde{H})$ by $\tilde{F}_{\cal
P}(\tilde{H})$ also modifies the normalization factor of
$\tilde{W}_{\cal P}$ but that is irrelevant for what follows.

\section{Asymptotic Form of $\tilde{F}_{\cal P}(\tilde{H})$ for $N
\gg 1$}
\label{asym}

For $N \gg 1$ we now display explicitly the dependence of
$\tilde{F}_{\cal P}(\tilde{H})$ on the normalized unitary invariants
$(1 / N) \langle \tilde{H}^n / \lambda \rangle$ with positive integer
$n$. We mention in passing that the terms of leading order in a
systematic expansion of $\tilde{F}_{\cal P}(\tilde{H})$ in powers of
$N_Q / N^2$ can also be obtained from the Harish--Chandra Itzykson
Zuber integral~\cite{Itz80,Zin03}, or from the standard supersymmetry
approach~\cite{Efe97,Ver85}. We use the assumption~(\ref{traceless})
and discuss the case where not all constraining matrices are traceless
in the Appendix. Then $\langle B(t) \rangle = 0$. We consider the
expressions
\be
\int {\rm d} [U] \ ( i \langle (\tilde{H} / \lambda) | U B(t) U^\dag
\rangle )^k
\label{B4}
\ee
with $k$ positive integer. Such expressions are generated when the
exponential in Eq.~(\ref{F7a}) is expanded in a Taylor series. To
calculate the integral over the unitary group we use a method valid
for $N \gg 1$~\cite{Pro02,Bro96}. To leading order in $1 / N$ the
integral can be done using Wick contraction on the matrices $U$, the
rules being
\be
U_{\mu \nu} U^\dag_{\rho \sigma} \to (1 / N) \delta_{\mu \sigma}
\delta_{\nu \rho} \ {\rm and} \  U_{\mu \nu} U_{\rho \sigma} \to 0 \ . 
\label{B5}
\ee
Terms of higher order are obtained in a similar fashion and lead to
similar results but are not considered here. We first look at a few
simple cases. For $k = 1$ the expression~(\ref{B4}) vanishes. For $k =
2$ and $k = 3$ we obtain $( i^2 / N) [ (1 / N) \langle (\tilde{H} /
\lambda)^2 \rangle ]$ $\langle B^2(t) \rangle$ and $2! (i^3 / N)^2
[ (1 / N) \langle (\tilde{H} / \lambda)^3 \rangle ] \langle B^3(t)
\rangle$, respectively. For $k = 4$ Wick contraction generates two
terms. One is proportional to the square of the $k = 2$ term just
considered. The other is given by $3! ( i^4 / N^3) [ (1 / N) \langle
(\tilde{H} / \lambda)^4 \rangle ] \langle B^4(t) \rangle$. For $k =
5$, Wick contraction generates two types of terms: The product of the
$k = 2$ term and the $k = 3$ term, and a new term given by $4! ( i^5
/ N^4)$ $[ (1 / N) \langle (\tilde{H} / \lambda)^5 \rangle ]$ $\langle
B^5(t) \rangle$. 

For the general expression~(\ref{B4}) we consider all partitions of
$k$ into sets of positive integers $k_1, k_2, \ldots k_f$ greater than
unity such that $\sum_{i = 1}^f k_i = k$. To leading order in $1 / N$,
expression~(\ref{B4}) is given by the sum over all such partitions,
the contribution of each partition being $i^{k} {k \choose k_1} {k -
k_1 \choose k_2} \times \ldots \times {k - k_1 - \ldots - k_{f - 1}
\choose k_f} \prod_{i = 1}^f (k_i - 1)! ( 1 / N^{k_i - 1} ) [ (1 / N)
\langle (\tilde{H} / \lambda)^{k_i} \rangle ] \langle B^{k_i}(t)
\rangle$. The terms of higher order in $1 / N$ also involve products
of traces of powers of $\tilde{H} / \lambda$ and of traces of $B$, the
difference being that at least one trace of a power of $\tilde{H} /
\lambda$ is multiplied by at least two traces of powers of $B$ such
that the sum of the exponents of $B$ equals the exponent of $\tilde{H}
/ \lambda$. It is shown below that $\langle B^n \rangle$ and $\prod_i
\langle B^{n_i} \rangle$ with $\sum_i n_i = n$ are of the same order
in $N$ so the neglect of such terms is legitimate.

We conclude that to leading order in $1 / N$, the integral over the
unitary group in Eq.~(\ref{F7a}) is given by
\ba
&& \int {\rm d} [U] \ \exp \{ i \langle U B(t) U^\dag | (\tilde{H} /
\lambda) \rangle \} \nonumber \\
&& \qquad = \exp \{ \sum_{n \geq 2} (1 / n) (i^n / N^{n - 1}) [ (1 / N)
\langle (\tilde{H} / \lambda)^n \rangle ] \langle B^n(t) \rangle \} \ .
\label{B6}
\ea
For $N \gg 1$ the constraining function $\tilde{F}$ is then given by 
\ba
\tilde{F}_{\cal P}(\tilde{H}) &=& \bigg( \frac{\langle (\tilde{H})^2
\rangle}{2 \pi \lambda^2 N^2} \bigg)^{N_Q / 2} \nonumber \\
&\times& \int {\rm d} t \ \exp \{ \sum_{n \geq 2} (1 / n) (i^n / N^{n
- 1} ) [ (1 / N ) \langle (\tilde{H} / \lambda )^n \rangle] \langle
B^n(t) \rangle \} \ .
\label{B7}
\ea

It may seem that because of the factors $N^{n - 1}$ the terms of
higher order in $n$ in Eqs.~(\ref{B6}) and (\ref{B7}) can be
neglected. We now show that for $N_Q \sim N^2$ this is not the
case. In Eq.~(\ref{B7}) we replace the Cartesian integration variables
$t_q$ by polar coordinates $\{ r, \Omega \}$ in $N_Q$ dimensions where
\be
r^2 = \sum_q t^2_q
\label{B8}
\ee
and where $\Omega$ stands for the angular variables. We write
\be
B(t) = r B(\Omega)
\label{B9}
\ee
and since $\langle B_q | B_{q'} \rangle = \delta_{q q'}$ have
\be
\langle B^2(\Omega) \rangle = 1 \ .
\label{B10}
\ee
Let $b_\mu(\Omega)$ denote the
$N$ real eigenvalues of $B(\Omega)$. Then $\sum_\mu b^2_\mu(\Omega) =
1$ and $|b_\mu(\Omega)|
\leq 1$ for all $\mu = 1, \ldots, N$. For integer $n > 2$ this implies
that
\be
\langle B^n(\Omega) \rangle \leq 1. 
\label{B11}
\ee
This, incidentally, justifies the omission of terms of order $1 / N$
above and shows that $\langle B^n(\Omega) \rangle$ and $\langle
(\tilde{H} / \lambda)^n \rangle$ are characteristically different: The
first expression is (at most) of order unity while the second is of
order $N$. That is why we always carry the second expression in the
form $(1 / N) \langle (\tilde{H} / \lambda)^n \rangle$.

Using the transformation to polar coordinates we observe that the term
with $n = 2$ in Eq.~(\ref{B7}) gives a Gaussian integral in
$r$. Expanding the terms with $n > 2$ in the exponent in a Taylor
series we are led to consider radial integrals of the form
\be
\int {\rm d} r \ r^{N_Q - 1 + 2 k} \exp \{ - c r^2 \} \ .
\label{B12}
\ee
Here $c$ is a constant and $2 k$ must be even as otherwise the
integrals vanish. Compared to the leading term ($k = 0$) these
integrals yield a factor $(N_Q + 2k - 2) (N_Q + 2 k - 4) \times \ldots
\times N_Q$. (We assume for simplicity that $N_Q$ is even). If the
expansion of the exponential converges sufficiently rapidly so that
for $N_Q \sim N^2$ we need consider only terms with $k \ll N_Q$ then
every power of $r$ in the exponential in Eq.~(\ref{B7}) effectively
carries a factor $\sqrt{N_Q}$, and the series in $n$ proceeds
effectively in powers of $N_Q / N^2$. For $N_Q \sim N^2$ that factor
is of order unity.

While it is, thus, not permitted to terminate for $N_Q \sim N^2$ the
series in $n$ in Eq.~(\ref{B7}) with the first few terms, rapid
convergence of the Taylor expansion of the exponential around the
Gaussian form is assured by the following property of the matrix
$B(\Omega)$ defined in Eq.~(\ref{B9}). Each term in the Taylor
expansion of the right--hand side of Eq.~(\ref{B7}) around the
Gaussian form ($n = 2$) generates a factor of the form
\be
\int {\rm d} \Omega \ \prod_i \langle B^{k_i}(\Omega) \rangle \ {\rm
where} \ \sum_i k_i = {\rm even} = 2 k \ {\rm and \ where \ all} \ k_i
\geq 3 \ .
\label{B13}
\ee
That expression can also be written as
\be
\int {\rm d} \Omega \ \prod_i \sum_{\mu = 1}^N b^{k_i}_\mu(\Omega) \ .
\label{B14}
\ee
In magnitude, each eigenvalue $b_\mu(\Omega)$ is bounded by unity. It
is, therefore, safe to expect that on averaging over the
$N_Q$--dimensional unit sphere, each eigenvalue is of order $1 /
\sqrt{N}$ so that the expression in Eq.~(\ref{B14}) is of order $N^{-
k} \Omega(N_Q)$. Here $\Omega(N_Q)$ is the surface of the unit
sphere in $N_Q$ dimensions. That shows that only few terms in the
expansion are expected to contribute significantly even for $N_Q \sim
N^2$. The statement holds {\it a fortiori} for $N_Q \ll N^2$.

\section{Proof}
\label{proof}

To calculate the influence of the constraints in Eq.~(\ref{F8}) on the
spectrum for $N \gg 1$, we use the approach developed in
Ref.~\cite{Hac95} based on the supersymmetry
method~\cite{Efe97,Ver85}. We only sketch the essential steps, using
the definitions and notation of Refs.~\cite{Ver85} and
\cite{Hac95}. The average level density and all correlation functions
are obtained with the help of a generating functional $Z$ which is
written as
\be
Z = \int {\rm d} \Psi \big \langle  \exp \big \{ \frac{i}{2}
\Psi^{\dag} {\bf L}^{1/2} {\bf G} {\bf L}^{1/2} \Psi \big \} \big
\rangle_{H}
\label{C1}
\ee
where
\be
{\bf G} = {\bf H} - {\bf E} + {\bf M} \ .
\label{C1a}
\ee
The average over the ensemble is denoted by angular brackets with an
index $H$ while for the trace we continue to use angular brackets
without index as in Eq.~(\ref{scalar}). The symbol $\Psi$ stands for a
supervector the dimension of which depends on the particular
correlation function under study. The same is true of the matrices
${\bf H}$ (the Hamiltonian), ${\bf E}$ (the energy), and ${\bf
M}$. The matrix ${\bf M}$ is of order $O(N^{-1})$ and contains energy
differences, source terms and possible couplings to channels.
Differentiation with respect to the source terms generates the
particular correlation function of interest.

The invariance of CGUE under unitary transformations implies that the
integrand in Eq.~(\ref{C1}) depends upon $\Psi$ and $\Psi^{\dag}$ only
via the invariant form
\be
A_{\alpha \beta} = N^{-1} L^{1/2}_{\alpha \alpha} \sum_{\mu = 1}^N
\Psi_{\mu \alpha} \Psi^{\dag}_{\mu \beta} L^{1/2}_{\beta \beta} \ . 
\label{C2}
\ee
Here $\alpha$ amd $\beta$ are matrix indices in superspace while $\mu$
runs over the $N$ basis states of Hilbert space ${\cal H}$. We
introduce a supermatrix $\sigma$ with the same dimension and symmetry
properties as $A$ by writing $Z$ as an integral over a delta function,
\be
Z = \int {\rm d} \Psi \int {\rm d} \sigma \ \delta(\sigma - A) \big
\langle  \exp \big \{ \frac{i}{2} \Psi^{\dag} {\bf L}^{1/2} {\bf G}
{\bf L}^{1/2} \Psi \big \} \big \rangle_{H} \ .
\label{C3}
\ee
The delta function is replaced by its Fourier transform, and the
multiple Gaussian integral over the supervector $\Psi$ is performed to
yield
\be
Z = \int {\rm d} \sigma \int {\rm d} \tau \ \exp\big\{ \frac{i}{2} N \
{\rm trg} ( \tau \sigma) \big\} \big \langle \exp \big \{ - \frac{1}{2}
{\rm tr} \ {\rm trg}\ln [{\bf G} - \lambda \tau] \big \} \big
\rangle_{H} \ .
\label{C4}
\ee
The remaining superintegrations over $\tau$ and $\sigma$ are
eventually done for $N \to \infty$ with the help of the saddle--point
approximation. Prior to that step, the average over the ensemble is
performed using Eq.~(\ref{F8}). We first integrate over the unitary
group. To leading order in $N^{-1}$ we obtain
\ba
&& \big \langle \exp \big \{ - \frac{1}{2} {\rm tr} \ {\rm trg} \ln
[{\bf G} - \lambda \tau] \big \} \big \rangle_{H} = \big \langle \exp
\big \{ - \frac{1}{2} {\rm tr} \ {\rm trg} \ln {\bf D} \big \}
\nonumber \\
&& \qquad \qquad \times \exp \big\{ - \frac{1}{2} {\rm tr} \ {\rm trg}
\ln \big[ 1 + \frac{1}{N} {\rm tr} D^{-1} {\bf M} \big] \big \} \big
\rangle_{H} \ .   
\label{C5}
\ea
Here ${\bf D} = {\bf x} - {\bf E} - \lambda \tau$ is diagonal in
Hilbert space, and the angular brackets now stand for the remaining
integration over the eigenvalues $\{x_\mu\}$. Under inclusion of the
terms which arise from $\tilde{W}_{\cal P}$ in Eq.~(\ref{F8}), the
exponent of the integrand is given by
\ba
&& - \frac{1}{2} {\rm tr} \ {\rm trg} \ln {\bf D} - \frac{1}{2} {\rm
tr} \ {\rm trg} \ln \big[ 1 + \frac{1}{N} {\rm tr} {\bf D}^{-1} {\bf M}
\big] \nonumber \\
&& \qquad + 2 \sum_{\mu < \nu} \ln | x_\mu - x_\nu | - \frac{N}{2
\lambda^2} \sum_\mu x^2_\mu + \ln \tilde{F}_{\cal P}(\tilde{H}) \ .
\label{C6}
\ea
Following Ref.~\cite{Hac95}, we perform the eigenvalue integration
using the saddle--point approximation for $N \gg 1$. In
expression~(\ref{C6}), all terms in the first line are at most of
order $N$ while the first two terms in the second line are of order
$N^2$. Omitting the terms in the first line but keeping $\ln
\tilde{F}_{\cal P}(\tilde{H})$ (depending on the value of $N_Q$, that
function may or may not be of order $N^2$), we put the derivatives of
the resulting expression with respect to the $x_\mu$ equal to zero and
obtain the $N$ saddle--point equations
\be
x_\mu = \frac{2 \lambda^2}{N} \sum_{\sigma \neq \mu} \frac{1}{x_\mu
- x_\sigma} +  \frac{2 \lambda^2}{N} \frac{1}{\tilde{F}_{\cal P}
(\tilde{x})} \frac{\partial \tilde{F}_{\cal P}(\tilde{H})} {\partial
x_\mu} \ , \ \mu = 1, \ldots, N \ .
\label{C7}
\ee
Without the term $\ln \tilde{F}_{\cal P}(\tilde{H})$ in
expression~(\ref{C6}), the saddle--point equations would have taken the
standard GUE form
\be
x_\mu  = \frac{2 \lambda^2}{N} \sum_{\sigma \neq \mu} \frac{1}{x_\mu
- x_\sigma} \ .
\label{C8}
\ee
To prepare for the treatment of Eq.~(\ref{C7}) we recall how
Eqs.~(\ref{C8}) are solved in Ref.~\cite{Bre78}. The variables $x_\mu /
\lambda$ are replaced by a single dimensionless continuous variable,
$x_\mu / \lambda \to \ve$. The normalized level density $\rho(\ve)$
with $\int {\rm d} \ve \ \rho(\ve) = 1$ of the ensemble (which at this
point is an unknown function) is introduced, and Eq.~(\ref{C8}) is
written in the form
\be
\ve = 2 P\int {\rm d} \ve' \ \frac{\rho(\ve')}{\ve - \ve'} \ .
\label{C9}
\ee
Here $P\int$ stands for the principal--value integral. Eq.~(\ref{C9})
has an electrostatic analogue and can be solved using the theory of
analytic functions. The result is Wigner's semicirle law for
$\rho(\ve)$. The generating functional is subsequently taken at the
saddle--point values for the $x_\mu$. All summations over $x_\mu$ in
$Z$ are, thus, replaced by integrations over $\ve$ with $\rho(\ve)$ as
weight function.

In applying that same method to Eqs.~(\ref{C7}) we introduce the (yet
unknown) normalized average level density $\rho_{\cal P}(\ve)$ of the
constrained ensemble in the sum on the right--hand side of
Eq.~(\ref{C7}). We also have to implement the change of variables
$x_\mu / \lambda \to \ve$ in $\tilde{F}_{\cal P}(\tilde{H})$ and its
derivative. As for $\tilde{F}_{\cal P}(\tilde{H})$, this is done by
replacing everywhere in Eq.~(\ref{B7}) the term $(1 / N) \langle (H /
\lambda)^n \rangle$ by $\langle \ve^n \rangle = \int {\rm d} \ve \
\ve^n \rho_{\cal P}(\ve)$. The form of $\tilde{W}(H)$ in Eq.~(\ref{F8})
implies that $\rho_{\cal P}(\ve) = \rho_{\cal P}(-\ve)$ so that only
terms with $n$ even survive, and we obtain
\ba
\tilde{F}_{\cal P} &=& \bigg( \frac{\langle \ve^2 \rangle}{2 \pi
N} \bigg)^{N_Q / 2} \nonumber \\
&& \times \int {\rm d} t \ \exp \{ \sum_{n \geq 1} (1 / (2 n)) ((-1)^n
/ N^{2 n - 1} ) \langle \ve^{2 n} \rangle \langle B^{2 n}(t) \rangle
\} \ .
\label{C10}
\ea
This is a function of the unknown level density $\rho_{\cal P}$ with a
rapidly converging Taylor expansion around the Gaussian term ($n = 1$).
In the derivative of $\tilde{F}_{\cal P}$, we substitute $x_\mu /
\lambda \to \ve$ after differentiating with respect to $x_\mu$. We
obtain
\ba
\frac{\lambda}{N} \frac{\partial \tilde{F}_{\cal P}}{\partial x_\mu}
&=& \frac{N_Q \ve}{N^2 \langle \ve^2 \rangle} \tilde{F}_{\cal P} +
\bigg( \frac{\langle \ve^2 \rangle}{2 \pi N} \bigg)^{N_Q / 2}
\sum_{n \geq 1} \frac{(-)^n \ve^{2 n - 1}}{N^{2 n + 1}} \int {\rm
d} t \ \langle B^{2 n}(t) \rangle \nonumber \\
&& \times \exp \{ \sum_{n \geq 1} (1 / (2 n)) ((-1)^n / N^{2 n - 1} )
\langle \ve^{2 n} \rangle \langle B^{2 n}(t) \rangle \} \ .
\label{C11}
\ea
The right--hand side of Eq.~(\ref{C11}) is a polynomial of odd order
in $\ve$ with rapidly decreasing coefficients.

As a result, the saddle--point equations~(\ref{C7}) take the form
of Eq.~(\ref{C9}), with $\ve$ replaced by an odd--order polynomial in
$\ve$ with rapidly decreasing coefficients. These coefficients depend
on $\rho_{\cal P}(\ve)$; the solution of Eq.~(\ref{C11}) must,
therefore, proceed iteratively, with the GUE average level density as
a starting point for calculating $\langle \ve^n \rangle$. In
Ref.~\cite{Pap06} it was shown perturbatively that $\rho(E)$ and
$\rho_{\cal P}(E)$ differ by a term of order $N_Q / N^2$. Therefore,
we expect the two level densities to differ significantly when $N_Q
\sim N^2$.

Returning to Eq.~(\ref{C6}) we perform the integration over the
variables $x_\mu$ by taking their values at the saddle points. That
means, for instance, that we write
\be
\frac{\lambda}{N} \sum_\mu \frac{1}{x_\mu - E - \lambda \tau} \to
\int {\rm d} E' \frac{\rho_{\cal P}(E' / \lambda)}{E' - E - \lambda
\tau} \ .
\label{C12}
\ee
This is the essential step: The summations over the eigenvalues
$x_\mu$ disappear in all expressions in the integrand of $Z$. Each
such summation is replaced by an energy integral involving the level
density $\rho_{\cal P}(\ve)$ of the constrained ensemble. This is the
only place where the constraints show up in the calculation. From here
on the calculation of the correlation functions for the CGUE and that
for the GUE run completely in parallel~\cite{Hac95}. It follows that
all correlation functions of the CGUE have the same form as their GUE
counterparts except that we have to replace the local average level
spacing of the GUE by that of the CGUE. This proves our theorem.

\section{Discussion}
\label{dis}

Our theorem holds in the limit $N \gg 1$ and for $N_Q < N^{\rm
crit}_Q$. The average level density of the CGUE may differ from that
of the GUE but all correlation functions have the same form for both
ensembles. Although our result is perhaps expected, to the best of our
knowledge this is the first time that GUE--type statistics has been
analytically proved for a class of ensembles different from the GUE.
Deviations of order $1 / N$ from the asymptotic form of the GUE
statistics exist, of course, even for the pure GUE and are expected
{\it a fortiori} for the CGUE. Our proof specifically applies to the
case of unitary invariance. We believe, however, that a corresponding
result holds also for the other symmetries.

The proof of the theorem rests on the fact that in the limit $N \gg 1$
and for $N_Q < N^{\rm crit}_Q$, the constraining function
$\tilde{F}_{\rm P}(\tilde{H})$ is free of singularities. The proof
holds independently of any specific properties of the constraining
matrices $B_q$. What happens for $N \gg 1$ but $N_Q \geq N^{\rm
crit}_Q$? That seems to depend on specific properties of the
constraints which determine the eigenvalues $b_\mu(s)$ and, thus, the
convergence properties of the integrals over $s$. Therefore, generic
statements about the spectral fluctuation properties of the CGUE
probably cannot be made for $N_Q \geq N^{\rm crit}_Q$.

One may speculate that with $N_Q$ increasing beyond the value $N^{\rm
crit}_Q$, the spectral fluctuations of the CGUE remain GUE--like until
$N_P = N^2 - N_Q$ is reduced to the value $N_P = N$ (where $H_P =
\sum_p h_p B_p$ may be a linear combination of $N$ commuting matrices
and, thus, integrable). But that speculation is surely incorrect.
Indeed, random band matrices with a band width less than or of order
$\sqrt{N}$ are known~\cite{Cas90,Fyo91} to possess localized
eigenfunctions and a Poisson spectrum. For such matrices, the number
$N_Q$ of constraints is at least of order $N^2 - N \sqrt{N}$ and, for
$N \gg 1$, obviously much larger than $N_Q$ but still much smaller
than $N^2 - N$.

It is of interest to discuss the embedded random $k$--body ensembles
EGUE($k$) (see Ref.~\cite{Mon75} and the review~\cite{Ben03}) in the
light of these considerations. The EGUE(k) models a Fermionic
many--body system with $k$--body interactions: $m$ identical spinless
Fermions occupy $l$ degenerate single--particle states. The Hilbert
space is spanned by $N = {l \choose m}$ Slater determinants. To
construct the $k$--body interaction operators, we denote by
$a^\dag_\mu$ and $a^{}_\mu$ the creation and destruction operators for
a Fermion in the single--particle state labelled $\mu$ with $\mu = 1,
\ldots, l$. Let $\mu_1, \ldots, \mu_m$ with $1 \leq \mu_i \leq l$ for
all $i$ denote a set of $m$ non--equal integers, and analogously for
$\nu_1, \ldots, \nu_m$ with $1 \leq \nu_j \leq l$ for all $j$. Then a
general interaction operator has the form $A(\{ \mu_i \}, \{ \nu_j \})
= \prod_{i = 1}^m a^{\dag}_{\mu_i} \prod_{j = 1}^m a^{}_{\nu_j}$. In
the Hilbert space of Slater determinants, the $N^2$ Hermitean
operators $A(\{ \mu_i \}, \{ \nu_j \})$ $+$ $A^\dag(\{ \mu_i \}, \{
\nu_j \})$ and $i [ A( \{ \mu_i \}, \{ \nu_j \})$ $-$ $A^\dag( \{
\mu_i \}, \{ \nu_j \}) ]$ play the very same role as do the matrices
$B_\alpha$ introduced in Section~\ref{def} in the Hilbert space ${\cal
H}$. From the general form of $A$, a $k$--body operator is obtained by
imposing the condition that a subset of $m - k$ elements of the set
$\{ \mu_i \}$ is identically equal to a subset of $m - k$ elements of
the set $\{ \nu_j \}$. There are ${l \choose m} {m \choose k} {l - m
\choose k}$ such $k$--body operators. The EGUE($k$) is obtained by
writing the Hamiltonian as a linear combination of all $k'$--body
operators with $k' \leq k$ and with coefficients that are real random
Gaussian--distributed variables.

Among the EGUE($k$), the EGUE($2$) has received particular attention
because it mimics a Hamiltionian with two--body interactions, a form
typical for Fermionic many--body systems like atoms or nuclei. One of
the central questions (undecided so far) has been whether for $N \gg
1$ the spectral fluctuation properties of EGUE($2$) are GUE--like.
Numerical simulations~\cite{Ben03} suggest that the answer is
affirmative. However, these simulations are typically done for small
values of $l$ and $m$, with $l$ around $12$ and $m$ around $4$ or
so. But for these values the number of one-- plus two--body
interaction terms is ${12 \choose 4} [ 4 \times 8 + 6 \times 28 ] =
200 {12 \choose 4}$. The number of constraints is accordingly given by
$N_Q = {12 \choose 4} [ 495 - 200 ] = {12 \choose 4} \times 295$. That
figure is not much larger than $N^{\rm crit}_Q = {12 \choose 4} \times
247$, so that it is difficult to draw firm conclusions. It would be
more informative to investigate numerically large values of $l$ and
$m$ but that is prohibitively difficult. For $l \gg m \gg 1$ we have
$N \approx l^m$, and the number of independent two--body operators is
approximately $l^m m^2 l^2$. In other words, there are only $m^2 l^2$
non--zero interaction matrix elements in every row and column of the
matrix representation of the Hamiltonian for the EGUE($2$), much fewer
than for a banded random matrix where that number would be
approximately $\sqrt{N} = l^{m / 2}$. Put differently, the number
$N_Q$ of constraints for the EGUE($2$) is much bigger than it is for a
banded random matrix. That fact suggests that mixing in the EGUE($2$)
is weaker than it is for a banded random matrix, and that EGUE($2$)
has Poissonian level statistics. On the other hand, in a banded random
matrix it takes approximately $\sqrt{N}$ different interaction matrix
elements to connect two arbitrary states in Hilbert space. In the
EGUE($2$) that number is only $m / 2$, i.e., less even than $(1 / 2)
\ln N$. This fact suggests that mixing of the states in Hilbert space
is much more efficient for the EGUE($2$) than it is for a banded
random matrix, and the question remains undecided. But the discussion
suggests that for $N_Q \geq N^{\rm crit}_Q$, the form of the
constraints (and not just their sheer number) becomes important in
determining the spectral fluctuation properties of CGUE.

Another frequently used ensemble that simulates the nuclear many--body
system is the two--body random ensemble (TBRE), see
Refs.~\cite{Fre70,Boh71} and the review~\cite{Pap07}. Actually that
ensemble is taken to be invariant under time reversal and, thus, has
orthogonal rather than unitary symmetry. For simplicity we disregard
this fact. The single--particle states belonging to a major shell of
the nuclear shell model are occupied by a number of nucleons.  The
resulting Slater determinants are coupled to states with fixed total
spin $J$ and isospin $T$ and are written as $| J T \mu \rangle$.  The
running index $\mu$ has a typical range $R$ from several ten ($J$
large) to several thousand or more ($J$ small). Level statistics can
be meaningfully discussed only for large $R$. It is assumed that the
interaction between nucleons is of two--body type. Within a major
shell, the number of independent two--body matrix elements $v_\alpha$
is small (of order $10$ or $10^2$) compared to the large values of $R$
that are of interest. These matrix elements are taken to be
uncorrelated Gaussian--distributed random variables. This defines the
TBRE. For a given set of states $| J T \mu \rangle$, the matrix
representation of the Hamiltonian $H_{\rm TB}$ of the TBRE takes the
form \be (H_{\rm TB})_{\mu \nu} = \sum_\alpha v_\alpha C^{J T}_{\mu
\nu}(\alpha) \ .
\label{E1}
\ee
The matrices $C^{J T}_{\mu \nu}(\alpha)$ are fixed by the major shell
and by the quantum numbers $J$ and $T$ under consideration but have
some properties in common with matrices drawn from a canonical
random--matrix ensemble. Again, it is of interest whether in the limit
of $R \gg 1$ the TBRE generically obeys GUE (or GOE) level
statistics. Numerical results and semi--analytical arguments both
support such a hypothesis. Unfortunately, the matrices $C^{J T}_{\mu
\nu}(\alpha)$ are not accessible analytically so far. Therefore, it
does not even seem possible to determine the number $N_Q$ of
constraints that would characterize the TBRE matrix~(\ref{E1}), and we
cannot apply our results to that ensemble.

The authors of Ref.~\cite{Pap06} considered not only the CGUE but in
addition also what they called ``Deformed Gaussian Ensembles''. Here
the delta functions in Eq.~(\ref{A9a}) are replaced by Gaussians, and
the constraining function $F_{\cal P}(H)$ is everywhere regular.
Following the arguments in Section~\ref{proof} we conclude that the
spectral fluctuations of the deformed ensembles coincide with those of
the GUE.  In other words, constraints affect the spectral fluctuation
properties only if they constrain the relevant matrix elements to the
value zero (and not to very small non--zero values). As a consequence,
in the GUE the transition from GUE to Poisson level statistics is not
a continuous process (where the level statistics would be smoothly
deformed) but actually happens discontinuously. These statements apply
on the ``macroscopic'' level where the values of the coefficients
$h_q$ of the constraining matrices $B_q$ are compared with those of
the $h_p$ multiplying $B_p$. If, on the other hand, the $h_q$ are
measured in units of the mean level spacing (i.e., in effect, on a
scale $1 / \sqrt{N}$ compared to the scale of the $h_p$) then the
transition from GUE to Poisson level statistics is expected to be
smooth and to allow for intermediate forms of the level
statistics. That expectation is supported by transitions between
symmetries like the GOE $\to$ GUE transition, and by many examples of
partially chaotic systems that show intermediate level statistics. We
have not attempted to introduce a correspondingly scaled
parametrization for the deformed ensembles. Such a step would be
meaningful only in the immediate vicinity of the transition point from
GUE to Poisson level statistics. That point is not known analytically,
however.

\section*{Acknowledgments}
 
ZP thanks the members of the Max-Planck-Institut f{\"u}r Kernphysik in
Heidelberg for their hospitality and support, acknowledges support
by the Czech Science Foundation in Prague under grant no. 202/09/0084,
and thanks J. Kvasil and P. Cejnar for stimulating discussions.

\section*{Appendix: Constraining matrices with non--zero traces}

It is convenient to relabel the indices $q$ so that they run from $1$
to $N_Q$. We use Eq.~(\ref{F1}) and rotate the basis $B_q \to
\tilde{B}_q = \sum_{q'} O_{q q' } B_{q'}$ with the help of an orthogonal
transformation $O_{q q'}$ such that $\tilde{B}_1$ points in the
direction of the unit matrix ${\bf 1}_N$. Then,
\ba
\langle \tilde{B}_q | \tilde{B}_{q'} \rangle &=& \delta_{q q'} \ {\rm
for \ all} \ q,q' \ , \nonumber \\
\langle \tilde{B}_q \rangle &=& 0 \ {\rm for \ all} \ q > 1
\ .
\label{AB2}
\ea
We apply the same orthogonal transformation to the variables $t_q$ so
that $t_q \to \tilde{t}_q = \sum_{q'} O_{q q' } t_{q'}$ and obtain
\be
F_{\cal P}(H) = \bigg( \frac{\lambda^2}{2 \pi N} \bigg)^{N_Q/2} \int
\prod_q {\rm d} \tilde{t}_q \int {\rm d} [U] \ \exp \big( i \sum_q
\tilde{t}_q \langle \tilde{B}_q | U H U^{\dag} \rangle \big) \ .
\label{AB1a}
\ee
We write $\tilde{B}_1$ as the sum of a traceless part and of a multiple
of the unit matrix,
\be
\tilde{B}_1 = \bigg( \tilde{B}_1 - \langle \tilde{B}_1 \rangle \frac{{\bf
1}_N}{N} \bigg) + \langle \tilde{B}_1 \rangle \frac{{\bf 1}_N}{N} \ .
\label{AB3}
\ee
By construction, the traceless part is orthogonal to all the
$\tilde{B}_q$'s with $q > 1$ and has norm
\be
\langle \tilde{B}_1 - \langle \tilde{B}_1 \rangle \frac{{\bf 1}_N}{N} |
\tilde{B}_1 - \langle \tilde{B}_1 \rangle \frac{{\bf 1}_N}{N} \rangle =
1 - \frac{1}{N} \langle \tilde{B}_1 \rangle^2 =  \alpha^2 \ . 
\label{AB4}
\ee
Because of the first of Eqs.~(\ref{AB2}), the eigenvalues $b_{1 \mu}$
with $\mu = 1, \ldots, N$ of $\tilde{B}_1$ obey $\sum_\mu b^2_{1 \mu}
= 1$. We maximize $\langle \tilde{B}_1 \rangle= \sum_\mu b_{1 \mu}$
under that constraint and find that
\be
- \sqrt{N} \leq \langle \tilde{B}_1 \rangle \leq \sqrt{N} \ . 
\label{AB5}
\ee
Therefore,
\be
0 \leq \alpha^2 \leq 1 \ .
\label{AB6}
\ee
We define $\alpha$ as the positive root of $\alpha^2$ and write
Eq.~(\ref{AB3}) in the form
\be
\tilde{B}_1 = \alpha \hat{B}_1 + \langle \tilde{B}_1 \rangle
\frac{{\bf 1}_N}{N} \ .
\label{AB3a}
\ee
Then, $\hat{B}_1$ has trace zero and norm one. We define $\hat{B}_q =
\tilde{B}_q$ for all $q > 1$ and have
\be
\langle \hat{B}_q \rangle = 0 \ {\rm for \ all} \ q \ {\rm and} \
\langle \hat{B}_q | \hat{B}_{q'} \rangle = \delta_{q q'} \ {\rm for
\ all} \ q,q' \ .
\label{AB7}
\ee
For $\alpha = 0$ the matrix $\tilde{B}_1$ is a multiple of the unit
matrix, and the integral over $\tilde{t}_1$ in Eq.~(\ref{AB1a}) yields
a multiple of the delta function for $\langle H \rangle$ while the
remaining $N_Q - 1$ integrations over the $\tilde{t}_q$ with $q > 1$
are treated as in Section~\ref{asym}. For $\alpha = 1$ the matrix
$\tilde{B}_1$ is actually traceless; that case was treated in
Section~\ref{asym}. Therefore, we consider $\alpha$ only in the open
interval
\be
0 < \alpha < 1 \ .
\label{AB6a}
\ee
We rewrite Eq.~(\ref{AB1a}) by using the decomposition~(\ref{AB3a}), by
rescaling $\alpha \tilde{t}_1 \to \tilde{t}_1$, by introducing
spherical polar coordinates $\{ t, \Omega\}$ in $N_Q$ dimensions, and
by defining the matrix $B(\Omega)$ as
\be
t B(\Omega) = \sum_q \tilde{t}_q \hat{B}_q \ .
\label{AB8}
\ee
The function $F_{\cal P}(H)$ takes the form
\ba
F_{\cal P}(H) &=& \bigg( \frac{\lambda^2}{2 \pi N} \bigg)^{N_Q / 2}
\frac{1}{\alpha} \int {\rm d} t \ t^{N_Q - 1} {\rm d} \Omega \exp \big(
i \tilde{t}_1(t, \Omega) \langle \tilde{B}_1 \rangle \langle H \rangle /
(\alpha N) \big) \nonumber \\
&& \times \int {\rm d} [U] \ \exp \big( i t \langle B(\Omega) | U H
U^{\dag} \rangle \big) \label{AB1b} \ .
\ea
Here $\tilde{t}_1(t, \Omega)$ stands for the rescaled old integration
variable $\tilde{t}_1$ as expressed in terms of the new integration
variables $\{t, \Omega\}$. The function $F_{\cal P}(H)$ diverges for
$\langle H \rangle \to 0$. The divergence is removed by multiplying
$F_{\cal P}(H)$ with $[\langle H / \lambda \rangle^2]^{N_Q / 2}$.
As in the case of the substitution $F_{\cal P} \to \tilde{F}_{\cal P}$
in Section~\ref{mod}, we expect that this step does not affect the
spectral fluctuations of the ensemble. The matrix $B(\Omega)$ is
traceless by construction, and the integration over the unitary group
can be carried out as in Section~\ref{asym}. From here on we proceed
as in Section~\ref{proof}.

\end{document}